# Hyperbolic Shear Metasurfaces

Enrico M. Renzi[1,2], Emanuele Galiffi[1], Xiang Ni[1,3], Andrea Alù[1,2*]

[1]*Photonics Initiative, Advanced Science Research Center, City University of New York, New York, NY 10031, USA*
[2]*Physics Program, The Graduate Center, City University of New York, New York, NY 10026, USA*
[3]*School of Physics, Central South University, Changsha, Hunan, 410083, China*
[*]To whom correspondence should be addressed (email: aalu@gc.cuny.edu)

Polar dielectrics with low crystal symmetry and sharp phonon resonances can support hyperbolic shear polaritons – highly confined surface modes with frequency-dependent optical axes and asymmetric dissipation features. So far, these modes have been observed only in bulk natural materials at mid-infrared frequencies, with properties limited by available crystal geometries and phonon resonance strength. Here we introduce hyperbolic shear metasurfaces: ultrathin engineered surfaces supporting hyperbolic surface modes with symmetry-tailored axial dispersion and loss redistribution that can maximally enhance light-matter interactions. By engineering effective shear phenomena in these engineered surfaces, we demonstrate geometry-controlled, ultra-confined, low-loss hyperbolic surface waves with broadband Purcell enhancements, applicable across a broad range of the electromagnetic spectrum.

Hyperbolic waves emerge in materials featuring extreme optical anisotropy, with opposite signs of the real part of permittivity for orthogonal orientations of the electric field. These waves offer a powerful platform for nanophotonics, thanks to the open topology of their dispersion contours, which asymptotically stretch in momentum space, enabling subdiffractional light confinement combined with directional, ray-like propagation [1-7]. In turn, these features enhance the spontaneous emission rate of localized optical sources over broad bandwidths. While hyperbolic wave propagation has been originally explored in the context of metamaterials [3,7,8], it can also be found in natural materials, in particular in polar van der Waals dielectrics [9-12] and other low-symmetry polar crystals [13-17], which naturally exhibit extreme optical anisotropy associated with directional phonon resonances, leading to hyperbolic phonon polaritons. These hybrid light-matter quasiparticles arise within the Reststrahlen frequency band, within which one component of the permittivity tensor is negative, while the orthogonal one remains positive. While appealing because of their broad availability and lack of nanofabrication requirements, natural hyperbolic materials are restricted to specific frequency ranges in the mid-infrared regime [12,14,16].

Whether natural or engineered, bulk hyperbolic waves suffer from dissipation, and their light-matter interactions are hindered by material loss or metamaterial granularity [18]. By contrast, *hyperbolic metasurfaces* [19,20], characterized by highly directional in-plane resonances combined with subwavelength thickness, support effective Reststrahlen bands for surface waves, leading to a reduced impact of material loss and an easier access, since these modes live at the interface with air. These metasurfaces have been implemented across a wide range of frequencies, with exciting prospects for enhanced surface-wave manipulation and broadband interactions with localized emitters close to the surface [11,19,21-23,25].

Recently, a new family of bulk and surface hyperbolic phonon polaritons was unveiled in monoclinic polar crystals, known as *hyperbolic shear phonon polaritons* [14,15,17]. The non-orthorhombic lattice is associated with directional detuned resonances that are not orthogonal. Within the Reststrahlen band, hyperbolic modes can emerge also in this skewed lattice, but in contrast with conventional hyperbolic polaritons these modes feature a peculiar rotation of their optical axis with frequency (axial dispersion) and an asymmetric distribution of losses in the different branches of the hyperbolic iso-frequency contours (IFCs), driven by microscopic shear phenomena. These features endow hyperbolic shear polaritons with even stronger directionality and field confinement than conventional hyperbolic waves. In turn, shear phonon polaritons are only available in a limited set of natural materials with low crystal symmetry. Consequently, their observation has so far been limited to a non-optimal subset of possible lattice symmetries and phonon responses.

Here, we introduce and explore *hyperbolic shear metasurfaces*, structured surfaces engineered to support hyperbolic surface modes experiencing effective shear phenomena. By optimally breaking the in-plane symmetry of the metasurface, we induce strong axial dispersion and loss redistribution. In turn, this enables directional wave propagation stemming from intrinsic broken symmetry of the modal dispersion, rather than being induced by the excitation [7,26], or by slanted boundaries [27,28]. We show that the shear response can be tuned and optimized by rotating the relative angle between detuned directional in-plane resonances, leading to enhanced light-matter interactions and exotic propagation features compared to hyperbolic metasurfaces composed of the same underlying elements, but in a higher-symmetry configuration.

Consider a metasurface formed by a subwavelength array of detuned dipolar resonators [Fig. 1], placed in free space on the $z=0$ plane. In the long-wavelength limit, the optical response can be described by the $2\times2$ homogenized sheet conductivity tensor

$$\hat{\sigma} = \begin{pmatrix} \sigma_1 + \sigma_2 \cos^2(\theta) & -\sigma_2 \sin(\theta)\cos(\theta) \\ -\sigma_2 \sin(\theta)\cos(\theta) & \sigma_2 \sin^2(\theta) \end{pmatrix}, \tag{1}$$

where $\sigma_1$ and $\sigma_2$ are the effective complex conductivities associated with the two sets of dipolar resonators $R_1$ and $R_2$, which in Eq. (1) are assumed to be respectively oriented along the $x$-axis and at an angle $\theta$ from it, such that $\hat{\sigma} = \sigma_1 \hat{x} \otimes \hat{x} + \sigma_2 \hat{R}(\theta)(\hat{x} \otimes \hat{x})(\theta)\hat{R}^{-1}$, where $\hat{R}$ is a rotation matrix. The scalar response of each resonator is described by a Lorentzian dispersion $\sigma_j(\omega) = i\sigma_0 N_j \omega^2 / (\omega^2 - \Omega_j^2 + i\omega\Gamma_j)$, where $j = 1,2$, consistent with the dispersion of polaritonic crystals [16]. Without loss of generality, we assume that the two resonators feature equal loss rates $\Gamma_2 = \Gamma_1$, detuned resonance frequencies $\Omega_1$ and $\Omega_2 = 2\Omega_1$, and oscillator strengths $N_1$ and $N_2 = N_1/2$. The propagation of surface waves over such an impedance sheet obeys the dispersion relation [29]

$$k_x^2 \sigma_1 + \left(k_x \cos\theta - k_y \sin\theta\right)^2 \sigma_2 - k_0^2(\sigma_1 + \sigma_2) = 2k_0 k_z \left(1 + \frac{1}{4}\sigma_1\sigma_2 \sin^2\theta\right). \quad (2)$$

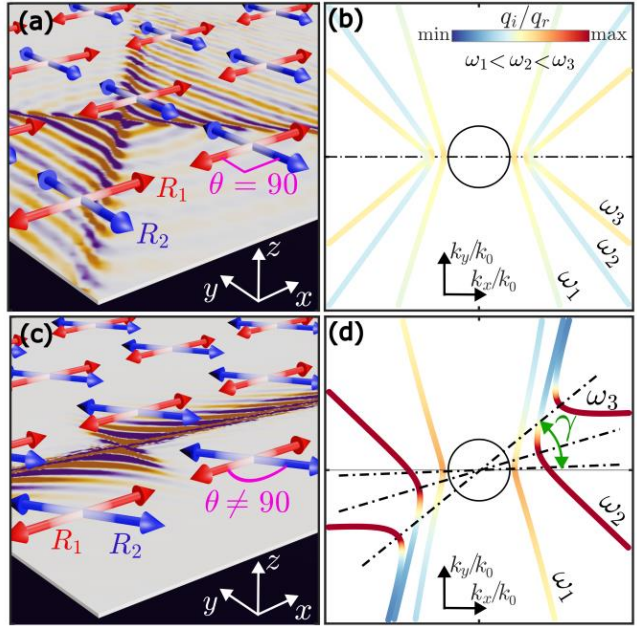

FIG. 1. Axial dispersion and loss redistribution in hyperbolic shear metasurfaces. (a) A metasurface composed of orthogonal detuned resonators supporting (b) hybrid-TM surface modes with hyperbolic IFCs, featuring non-dispersive optical axes (dash-dotted line) and symmetric damping (colormap). (c) By rotating the resonators, (d) the optical axis (dot-dashed lines) becomes frequency dispersive, featuring a rotation by an angle $\gamma(\omega,\theta)$ (green) relative to the $k_x$ axis, and the loss becomes asymmetric between the arms of the hyperbolic IFCs (calculated for $\theta = 60\,\mathrm{deg}$ and the angular frequencies $\Omega_1 < \omega_i < \Omega_2$, $i = 1,2,3$).

For orthogonal resonators, $\theta = 90\,\mathrm{deg}$ [Fig. 1(a)], the metasurface is uniaxial, described by a diagonal conductivity tensor $\mathrm{diag}(\sigma_1, \sigma_2)$. Hyperbolic surface waves are supported when $\Im[\sigma_1] > 0$ and $\Im[\sigma_2] < 0$ [19], within the effective Reststrahlen band of the homogenized surface $\Omega_1 < \omega < \Omega_2$. In this regime, the metasurface features tightly confined, highly directional hybrid transverse magnetic (TM) modes with hyperbolic IFCs. The aperture angle of the hyperbolas varies with frequency, while the optical axes are aligned with the orthogonal resonators [Fig. 1(b)]. The IFCs are highly symmetric, and even in the case of asymmetric damping rates in the two resonators ($\Re[\sigma_1] \neq 0 \neq \Re[\sigma_2]$) the four hyperbolic branches feature the same absorption features [19,21,25].

Effective shear phenomena are introduced by rotating $R_2$ with respect to $R_1$ [Fig. 1(c)]. In this scenario, $\hat{\sigma}$ acquires non-zero off-diagonal components, coupling the two polarization responses. Remarkably, this metasurface still supports hyperbolic surface waves in its Reststrahlen band, defined as the frequency range for which the two eigenvalues of the Hermitian part of $\hat{\sigma}$ have opposite signs. The symmetry axes of the hyperbolic IFCs are aligned with the reference system that diagonalizes this tensor at any given frequency. Given the non-orthogonality and detuning of the underlying resonators, the hyperbolas rotate with frequency, leading to *axial dispersion*, i.e., a frequency-dependent rotation of the IFCs by the angle $\gamma(\omega,\theta)$ [Fig. 1(d)]. This rotation is quantified by the angle subtended by the $k_x$-axis and the symmetry axis of the hyperbolic IFCs [14, 29]:

$$\gamma(\omega,\theta) = -\frac{1}{2}\tan^{-1}\left(\frac{\Im[\sigma_2]\sin 2\theta}{\Im[\sigma_1]-\Im[\sigma_2]\cos(2\theta)}\right). \tag{3}$$

Moreover, in contrast to the orthogonal scenario, the loss is distributed asymmetrically across the optical axes. This is to be expected, since loss is associated with the resonator polarization directions, and in this non-orthogonal scenario the electric field polarizations more aligned with the resonators (on two of the four hyperbolic branches) experience larger losses than the other two, which support electric fields less parallel to the resonators and hence less impacted by dissipation. In Fig. 1(d), we quantify the damping factor $\eta = q_i/q_r$ for each point on the IFCs (colormap), where $q_r$ ($q_i$) is the real (imaginary) component of the in-plane wave vector $\mathbf{q} = \mathbf{q}_r + \mathbf{q}_i$ of the surface mode.

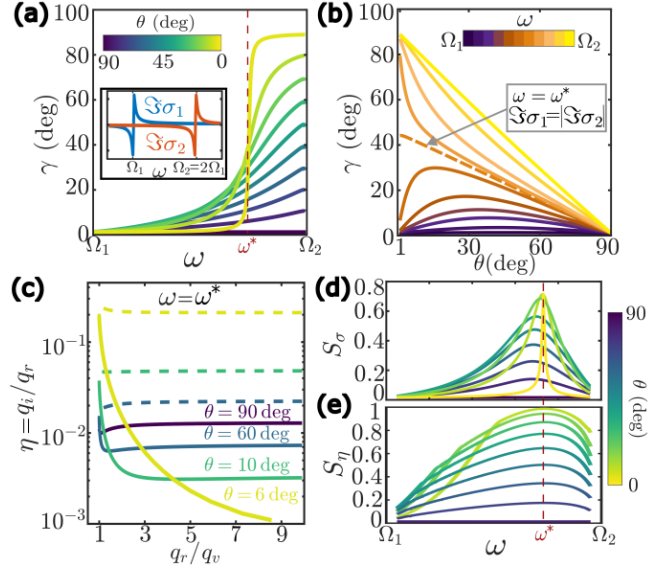

FIG. 2. Axial dispersion and loss redistribution. (a) Dispersion of the optical axis angle $\gamma(\omega,\theta)$. Small twist angles $\theta$ (colormap) narrow most of the dispersion around the critical frequency $\omega^*$, at which $\Im[\sigma_1] = -\Im[\sigma_2]$. The inset shows the dispersion of the reactive conductivity of $R_1$ and $R_2$ within the Reststrahlen band. (b) If $\omega < \omega^*$ (colormap), $R_1$ dominates, and $\gamma(\omega,\theta)$ reaches a maximum before turning back to zero as $\theta$ increases. For $\omega > \omega^*$, $R_2$ is dominant, and the rotation angle varies monotonically with $\theta$. (c) Loss redistribution between the hyperbolic arms as the rotation angle $\theta = (90,60,10,6)$ deg is varied. The damping factor $\eta$ increases in the lossier branches (dashed lines) and decreases in the longer-lived branches (solid lines) as the resonators becomes less orthogonal, stemming from the IFC vertices, located at $q_r = q_v$. Both figures of merit for the degree of loss asymmetry (d) $S_\sigma$ and (e) $S_\eta$ peak at the critical frequency $\omega^*$.

Fig. 2(a) shows the dispersion of $\gamma(\omega,\theta)$, which grows monotonically with frequency within the effective Reststrahlen band, going from being aligned to $R_1$ ($\gamma = 0$ at $\Omega_1$) to a larger value at $\Omega_2$. Its evolution corresponds to a counter-clockwise rotation of the contours in momentum space. As the resonators become less and less orthogonal, the axial dispersion in Fig. 2(a) steepens, most sharply in the proximity of the *critical frequency* $\omega^*$, for which $\Im[\sigma_1] = -\Im[\sigma_2]$ i.e., the two resonators support a combined resonance as they exhibit equal and opposite reactance, yielding extreme axial dispersion in the presence of effective shear. Fig. 2(b) explicitly plots $\gamma(\omega,\theta)$ at the critical frequency, demonstrating how the angle between $R_1$ and $R_2$ may be used to dramatically tailor the orientation of the hyperbolic axes. For frequencies closer to the two resonances, the rotation is dominated by one of the resonators. For $\omega < \omega^*$, this manifests in the appearance of a frequency-dependent and controllable turning point beyond which the rotation angle $\gamma(\omega,\theta)$ reverts to zero, bringing the symmetry axis of the IFC parallel to $R_1$ [29]. Interestingly, at the critical frequency (dashed orange line in Fig. 2(b)) $\gamma(\omega,\theta)$ varies linearly with the rotation angle, ranging between 0 and $\pi/4$. Finally, for $\omega > \omega^*$, $R_2$ dominates, and $\gamma(\omega,\theta)$ decreases monotonically from $\pi/2$ to 0 following the rotation of $R_2$ [orange to yellow lines in Fig. 2(b)].

The effective shear phenomena driving axial dispersion are also responsible for loss redistribution between the hyperbolic branches (Fig. 1d), increasing the absorption in two of them while enhancing propagation for the other two. This feature produces two surprising effects: given two resonator lattices $R_1$ and $R_2$, rotation-induced shear boosts light confinement and propagation length by lowering the impact of loss for two hyperbolic branches, and it enhances the overall directionality of hyperbolic wave propagation. These effects arise throughout the entire Reststrahlen band. To quantitatively explore shear-driven loss redistribution, in Fig. 2(c) we plot the damping factor $\eta = q_i / q_r$ as a function of the normalized in-plane momentum $q_r / q_v$, where $q_v(\omega^*, \theta)$ is the vertex of the hyperbolic IFC [see inset in Fig. 4(a)] at the critical frequency $\omega^*$. Dashed and solid lines respectively denote the high- and low-loss branches of the same hyperbolic IFC. For orthogonal resonators (purple line), the hyperbolas are symmetric, and the loss in the two resonators equally impacts states that are related to each other by inversion with respect to one of the optical axes. As $\theta$ decreases and the resonators become more parallel (light-blue to yellow lines), the fixed amount of loss in the resonators is heavily redistributed, making two of the branches increasingly lossier, while freeing the other two as we depart from the hyperbola vertex at $q_r = q_v$.

Losses are associated with the non-Hermitian (real) part of $\hat{\sigma}$, whose non-diagonal entries are generally nonzero even after rotating the reference frame by the angle $\gamma(\omega, \theta)$ that diagonalizes the Hermitian part. These off-diagonal non-Hermitian components of $\hat{\sigma}$ quantify the effective shear in the metasurface, whose macroscopic effect is loss redistribution. As a consequence, we can define $S_\sigma(\omega, \theta) = \Re[\hat{\sigma}'_{xy}] / \sqrt{\Re[\hat{\sigma}'_{xx}]^2 + \Re[\hat{\sigma}'_{yy}]^2}$ as a measure of the metasurface shear [30]: for $S_\sigma = 0$ we expect perfectly symmetric hyperbolas, corresponding to orthogonal resonators, while at the other extreme $S_\sigma = 1$ supports maximum loss asymmetry. Fig. 2(d) shows $S_\sigma(\omega, \theta)$ as a function of frequency and rotation angle, with increasing shear as the angle between resonators is reduced and as we approach the critical frequency.

The definition of $S_\sigma(\omega, \theta)$ does not refer to the specific wave propagation problem at hand, since it does not involve the boundary conditions, and it is therefore agnostic to the specific dispersion relation of the eigenmodes of interest. Its expression measures the general degree of 'electromagnetic asymmetry' of the metasurface, and provides an indication of the choice of parameters that maximize it. We can also introduce a quantitative measure of the modal asymmetry $S_\eta(\omega, \theta) = (\eta_+ - \eta_-) / (\eta_+ + \eta_-)$, shown in Fig. 2(e), which explicitly quantifies the loss redistribution by measuring the degree of loss asymmetry in the surface eigenmodes. We evaluate the damping factors $\eta_+$ and $\eta_-$ for mirror-symmetric in-plane momenta located on the bright and dark branches of the IFCs in the limit of large $q_r$, for which $q_i / q_r$ saturates [29]. Both metrics show a similar dependence on frequency and twist angle, and peak at the critical frequency for small twist angles, i.e, nearly parallel resonators.

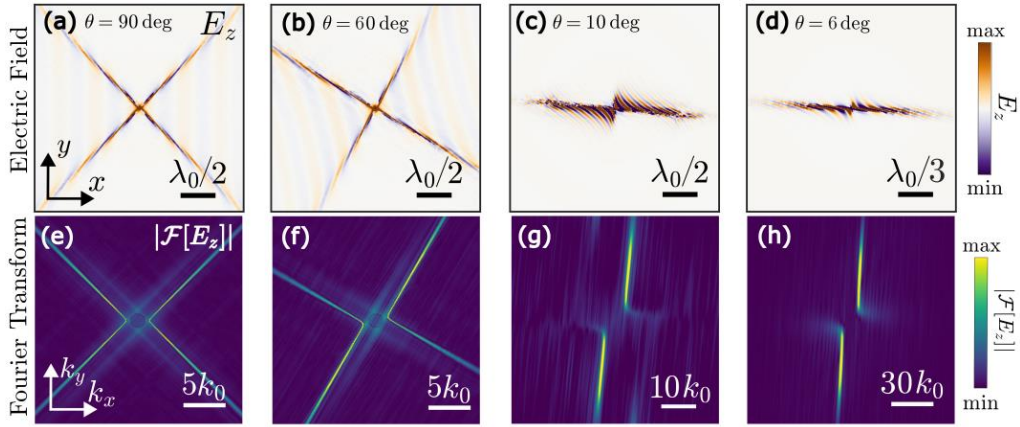

FIG. 3. Near-field excitation of hyperbolic shear surface waves by a localized emitter. Surface waves excited by a $z$-oriented electric point dipole placed at a distance $d_e = \lambda^* / 217$ from the metasurface. (a) For $\theta = 90 \deg$, normal electric field ($E_z$) and (e) its Fourier spectrum $|\mathcal{F}[E_z]|$ for excitation from a localized emitter. As the rotation angle between the same resonators varies (panels (b,c)), axial dispersion and asymmetric spectra emerge (panels (f,g)) showing the impact of shear. Remarkably, in (d,h) highly confined modes are observed for small rotation angles, exhibiting extraordinarily long-lived propagation enhanced by up to two orders of magnitde compared to orthogonal case, despite their tight confinement. Here, $\Omega_1 = \Omega_2 / 2 = 5$ GHz, $\gamma_1 = \gamma_2 = 0.02\Omega_1$, $N_1 = 2N_2 = 1$, and $\omega = \omega^* = 1.733\Omega_1$. These parameters match those experimentally used in recent metasurface experiments [23,31].

Fig. 3(a-d) and Fig. 3(e-h) show the spatial distribution of $E_z$ and the associated Fourier spectrum $|\mathcal{F}[E_z]|$ for surface waves launched by a $z$-oriented electric point dipole emitter placed above the metasurface at distance $\lambda^* / 217$ as we vary $\theta$. We show the results for the critical frequency $\omega^*$, at which the impact of effective shear is strongest. In the case of orthogonal resonators [Fig. 3(a,e)], the waves propagate symmetrically with respect to the optical axes, aligned with the resonators. As

the orthogonality is broken [Fig. 3(b,c)] the hyperbolic wavefronts rotate and the loss is redistributed, dampening two of the four ray-like beams. The lower loss branches experience further field confinement, as they access larger momenta [Fig. 3(c)]. By comparing Fig. 3(e) with Fig. 3(f,g), the extreme propagation asymmetry and enhanced directionality becomes apparent. We stress that in the different scenarios we are preserving the same underlying microstructure of the metasurface, i.e., the same features for $R_1$ and $R_2$, and the dramatic change in dispersion is only associated with the rotation between them.

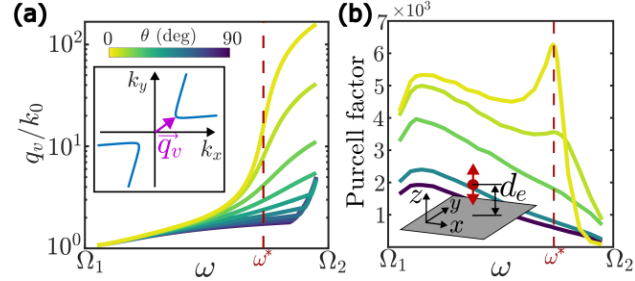

FIG. 4. Induced confinement and spontaneous emission rate (SER) enhancement for multiple angles between resonators. (a) Minimum wave number supported by the metasurface $q_v$ at the hyperbola vertex (purple arrow in the inset). (b) Mode confinement and increased lifetime produce a rotation-induced broadband SER enhancemnt for a z-oriented electric dipole emitter placed at distance $d_e = \lambda^*/217$ from the metasurface.

Fig. 4(a) shows the minimum momentum $q_v = |\vec{q}_v|$ of the hyperbolic IFCs, found at the IFC vertex (purple arrow in the inset), as we vary frequency and rotation angle $\theta$. As the frequency approaches $\Omega_2$ and $\theta$ decreases, deeply subdiffractional propagation is achieved - a by-product of effective shear (see [29] for further details). As a result of the lower loss and stronger field confinement, light-matter interactions are expected to be largely enhanced. Fig. 4(b) shows the enhancement of the Purcell factor [32,33] for a $z$-oriented electric point dipole placed at a distance $d_e = \lambda^*/217$ above the metasurface (inset). Compared to the orthogonal scenario (purple line), the broken symmetry produces a strong broadband enhancement of the emission rate (blue to yellow lines). This effect is consistent with the enhancement of $E_z$ in Fig. 3(a-d) as $\theta$ decreases. For small angles $\theta$ between the two resonators, the Purcell factor shows a broad enhancement across the entire Reststrahlen band, with a maximum at $\omega \approx \omega^*$.

In this Letter, we introduced hyperbolic shear metasurfaces which support effective shear phenomena for hyperbolic surface waves, induced by tailoring the angle between detuned directional resonances. Through these effects, we can tailor the directionality and boost the lifetime and Purcell factor of hyperbolic surface waves over tunable bandwidths, achieving extreme control over their propagation and dissipation features. Our work establishes a paradigm which leverages broken symmetries to realize low-loss, ultra-confined and highly directional surface wave propagation. The phenomena demonstrated may be applied to a wide range of frequencies. These metasurfaces may be realized at radio-frequency using twisted bi-layer of detuned resonators [23] or in optics using asymmetric V-shaped resonators meta-units [34]. The rational control achieved through the rotational degree of freedom implies that, through tailored optical pumps and nonlinearities, it may be possible to dynamically tune the effective shear in hyperbolic metasurfaces, thus enabling large tunability in real time. These tools may lead to time-dependent axial dispersion and loss redistribution, as well as opportunities for pulse shaping and multiplexing.

The authors thank Simon Yves for the fruitful discussions. This work was supported by the Office of Naval Research and the Simons Foundation. E.G. acknowledges funding from the Simons Foundation through a Junior Fellowship of the Simons Society of Fellows (855344, EG).

# Supplementary material for "Hyperbolic Shear Metasurfaces"

Enrico Maria Renzi[1,2], Emanuele Galiffi[1], Xiang Ni[1,3], Andrea Alù[1,2*]

[1]*Photonics Initiative, Advanced Science Research Center, City University of New York, New York, NY 10031, USA*

[2]*Physics Program, The Graduate Center, City University of New York, New York, NY 10026, USA*

[3]*School of Physics and Electronics, Central South University, Changsha, Hunan, 410083, China*

[*]To whom correspondence should be addressed (email: aalu@gc.cuny.edu)

# Contents

# 1 Dispersion equation for hyperbolic shear surface waves

## 1.1 Conductivity tensor for hyperbolic shear metasurface

We consider an infinitely thin metasurface in air placed on the $z=0$ plane, patterned with meta-units consisting of two detuned dipolar resonators $R_1$ and $R_2$, respectively. Resonators $R_1$ are aligned along the $x$-axis, while the latter are oriented at an angle $\theta$ which can be controlled arbitrarily.

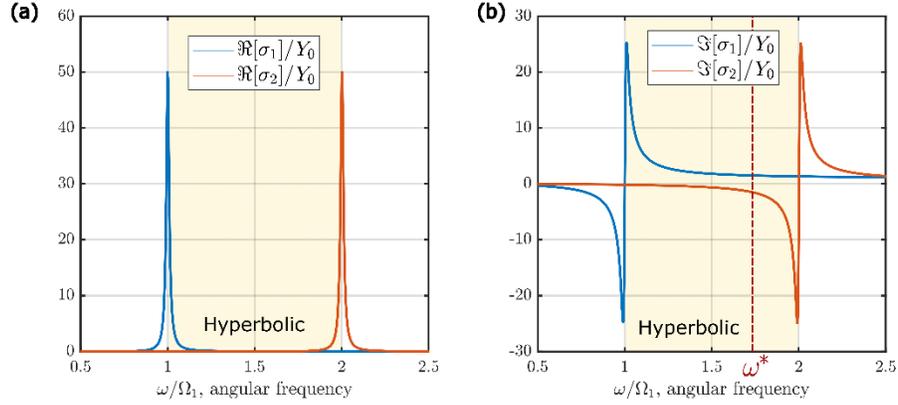

*Figure S1: (a) Normalized resistive part and (b) reactive part of $\sigma_1$ (blue) and $\sigma_2$ (orange), respectively, obtained using the Lorentzian parameters $\Omega_1 = \Omega_2/2 = 5$ GHz, $\gamma_1 = \gamma_2 = 0.02\Omega_1$ and $N_1 = 2N_2 = 1$ (the same parameters have been used in the main text). The dashed red line in plot (b) represents the critical frequency $\omega^*$, at which $\Im[\sigma_1] = \Im[\sigma_2]$. The hyperbolic region is highlighted in yellow, where $\Im[\sigma_1] > 0$ and $\Im[\sigma_2] < 0$.*

Defining $\sigma_1(\omega)$ and $\sigma_2(\omega)$ as the complex conductivities associated to the two sets $R_1$ and $R_2$, we may define the homogenized conductivity tensor for the sheet impedance as

$$\hat{\sigma} = \begin{pmatrix} \sigma_1 + \sigma_2 \cos^2(\theta) & -\sigma_2 \sin(\theta)\cos(\theta) \\ -\sigma_2 \sin(\theta)\cos(\theta) & \sigma_2 \sin^2(\theta) \end{pmatrix}, \tag{S.1}$$

where $\sigma_1(\omega)$ and $\sigma_2(\omega)$ are complex conductivities describing the strength of resonators $R_1$ and $R_2$, respectively. The conductivities $\sigma_1(\omega)$ and $\sigma_2(\omega)$ are Lorentzian functions (Fig. 1(a,b)) of the form [21]:

$$\sigma_j(\omega) = \frac{iY_0 N_j \omega^2}{\omega^2 - \Omega_j^2 + i\omega\Gamma_j}, \quad j=1,2, \tag{S.2}$$

which reproduce the typical response of phonon polaritons in Van der Waals materials, as pointed out in our letter. In Eq. (S.2), $Y_0$ is the free space conductivity, $N_j$ the resonator strength at infinite frequency, $\Omega_j$ the resonance frequency and $\Gamma_j$ the loss rate of the resonators (the adopted time convention is $e^{-i\omega t}$).

We define the critical frequency $\omega^*$ (red dashed line in Fig. 1(b)) as the frequency at which $\Im[\sigma_1] = \Im[\sigma_2]$ in the lossless scenario. From the definition in Eq. (S.2) we obtain:

$$\omega^* = \sqrt{\frac{N_1 \Omega_2^2 - N_2 \Omega_1^2}{N_1 - N_2}} \tag{S.3}$$

## 1.2 Eigenvalues of the conductivity tensor

Without loss of generality, we carry out the analysis assuming a lossless metasurface, i.e., setting $\Gamma_1 = 0 = \Gamma_2$ such that the conductivity tensor is $\hat{\sigma} = i\hat{\sigma}_i$, where $\hat{\sigma}_i \in \mathbb{R}^{2\times 2}$. Firstly, it is convenient to work in the frame of reference rotated by $\gamma(\omega, \theta)$, so that the reactive part of the conductivity tensor is diagonal, i.e.:

$$\hat{\sigma}' = \hat{R}_\gamma \hat{\sigma} \hat{R}_\gamma^T = i\hat{\sigma}_i' = i\begin{pmatrix} \sigma_{i,1}' & 0 \\ 0 & \sigma_{i,2}' \end{pmatrix}, \tag{S.4}$$

where $\sigma_{i,1}', \sigma_{i,2}' \in \mathbb{R}$ and $\hat{R}_\gamma = \begin{pmatrix} \cos\gamma & \sin\gamma \\ -\sin\gamma & \cos\gamma \end{pmatrix}$ is a rotation by the real angle $\gamma$. The corresponding eigenvalues of the matrix are

$$\sigma_{i,1}' = \frac{1}{2}\left(\sigma_{i,1} + \sigma_{i,2} + \sqrt{\sigma_{i,1}^2 + \sigma_{i,2}^2 + 2\sigma_{i,1}\sigma_{i,2}\cos(2\theta)}\right) \tag{S.5}$$

and

$$\sigma_{i,2}' = \frac{1}{2}\left(\sigma_{i,1} + \sigma_{i,2} - \sqrt{\sigma_{i,1}^2 + \sigma_{i,2}^2 + 2\sigma_{i,1}\sigma_{i,2}\cos(2\theta)}\right) \tag{S.6}$$

Interestingly, by construction $\sigma_{i,1}' > 0$ and $\sigma_{i,2}' < 0$, meaning that the metasurface remains hyperbolic within the entire Reststrahlen band (Fig. S2(a,b)). At the critical frequency the two eigenvalues are always equal in magnitude and opposite in sign for any value of the angle $\theta$ (dashed lines in Fig. S2(b)).

The condition $\theta = 0$ is the limiting case in which the two families of resonators are aligned. Below $\omega^*$ the metasurface is equivalent to an array of metal nano-antennas. At $\omega^*$, the metasurface behaves as an epsilon near-zero medium. Above $\omega^*$ the metasurface is equivalent to an array of dielectric dipoles and its response is analogous to a linear polarizer.

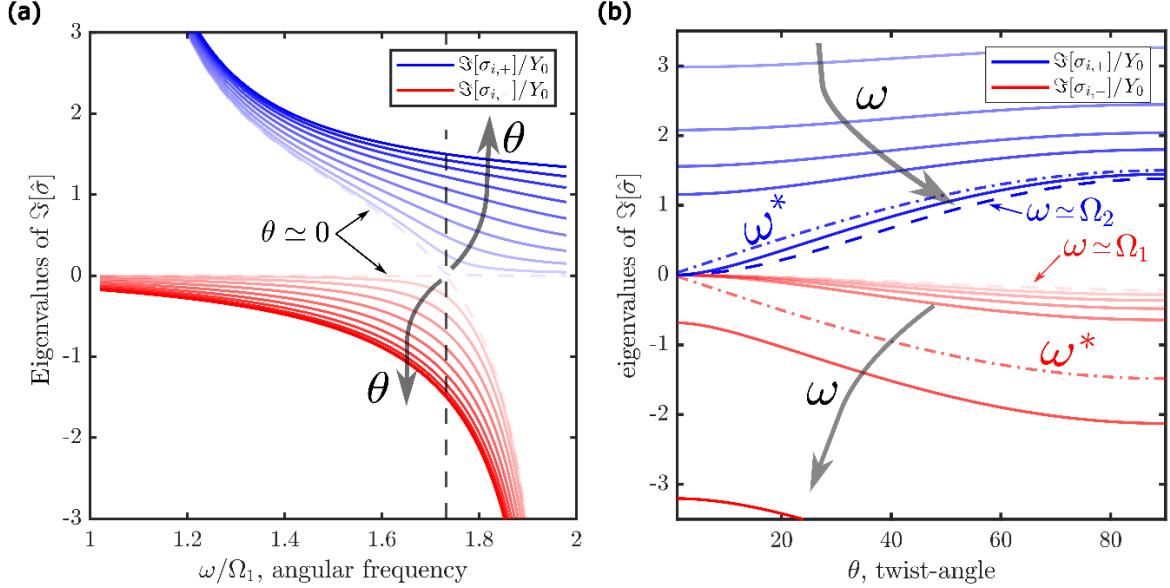

*Figure S2: (a) Across the entire Reststrahlen band $\Omega_1 < \omega < \Omega_2$, one eigenvalue of the reactivce part of the conductivity tensor is always positive ,i.e., $\Im[\sigma_{i,+}]$ (in blue), whereas the other stays negative, i.e. $\Im[\sigma_{i,-}]$ (in red.), for any value of the angle $\theta$ (increasing from the pale color to the vivid color). Therefore, the metasurface stays hyperbolic within the entire Reststrahlend band. In (b) we observe that if $\omega \simeq \Omega_1$ $\Im[\sigma_{i,+}] \gg |\Im[\sigma_{i,-}]|$, if $\omega = \omega^*$ one has $\Im[\sigma_{i,+}] = -\Im[\sigma_{i,-}]$, and, finally if $\omega \simeq \Omega_2$ $\Im[\sigma_{i,+}] \ll |\Im[\sigma_{i,-}]|$.*

### 1.3 Derivation of the dispersion equation

The dispersion equation is derived adopting the time convention $e^{-i\omega t}$. The single frequency electric and magnetic field of the surface waves guided by the metasurface can be written as:

$$\mathbf{E} = \mathbf{E}_0 e^{i\mathbf{k}\cdot\mathbf{r}}, \quad \text{and} \quad \mathbf{H} = \mathbf{H}_0 e^{i\mathbf{k}\cdot\mathbf{r}}, \qquad (S.7)$$

where the wave vector $\mathbf{k} = k_x\hat{\mathbf{x}} + k_y\hat{\mathbf{y}} + k_z\hat{\mathbf{z}}$, where $k_z = \sqrt{k_0^2 - k_x^2 - k_y^2}$, is defined for the general case of a lossy material, imposing the condition $\Re[k_z]\Im[k_z] \geq 0$, which guarantees the existence of propagating modes bound to the surface. Imposing the discontinuity of the magnetic field at $z = 0$, the boundary condition for the fields above the conductivity reads [35]:

$$\bar{\sigma}\cdot\mathbf{E}_{\|} = 2\hat{\mathbf{z}}\times\left(\mathbf{H}_{\|}\right)_{z=0}, \tag{S.8}$$

where the symbol $\|$ identifies the in-plane components of the electric field and magnetic field, respectively, and $\bar{\sigma} = \sigma_{xx}\hat{\mathbf{x}}\hat{\mathbf{x}} + \sigma_{xy}\hat{\mathbf{y}}\hat{\mathbf{x}} + \sigma_{yx}\hat{\mathbf{x}}\hat{\mathbf{y}} + \sigma_{yy}\hat{\mathbf{y}}\hat{\mathbf{y}}$ is the dyadic form of the permittivity tensor. Starting from Faraday law, we may write:

$$\mathbf{H}_0 = \frac{1}{\omega\mu_0}\mathbf{k}\times\mathbf{E}_0 \tag{S.9}$$

The cross product of both sides with $\hat{\mathbf{z}}$ gives

$$\hat{\mathbf{z}}\times\mathbf{H}_{0,\|} = \frac{1}{\omega\mu_0}\hat{\mathbf{z}}\times\mathbf{k}\times\mathbf{E}_0. \tag{S.10}$$

Using the vector calculus identity $\mathbf{a}\times(\mathbf{b}\times\mathbf{c}) = (\mathbf{a}\cdot\mathbf{c})\mathbf{b} - (\mathbf{a}\cdot\mathbf{b})\mathbf{c}$, we obtain

$$\hat{\mathbf{z}}\times\mathbf{H}_{0,\|} = \frac{1}{\omega\mu_0}\left[(\hat{\mathbf{z}}\cdot\mathbf{E}_0)\mathbf{k} - (\hat{\mathbf{z}}\cdot\mathbf{k})\mathbf{E}_0\right] \tag{S.11}$$

Substituting into Eq. (S.8), we obtain

$$\bar{\sigma}\cdot\mathbf{E}_{\|} = \frac{2}{\omega\mu_0}\left[(\hat{\mathbf{z}}\cdot\mathbf{E}_0)\mathbf{k} - (\hat{\mathbf{z}}\cdot\mathbf{k})\mathbf{E}_0\right] \tag{S.12}$$

Starting from Gauss law for the electric field, we can write the electric field z component $E_z$ in terms of its in-plane components $E_x$ and $E_y$ as

$$E_z = \left(\bar{I}_{\|} - \frac{1}{k_z}\hat{\mathbf{z}}\mathbf{k}_{\|}\right)\cdot\mathbf{E}_{0,\|}, \tag{S.13}$$

where $\bar{I}_{\|} = \hat{\mathbf{x}}\hat{\mathbf{x}} + \hat{\mathbf{y}}\hat{\mathbf{y}}$. Substituting into (S.12), we obtain

$$\bar{\sigma}\cdot\mathbf{E}_{\|} = \frac{2}{\omega\mu_0}\left[(\hat{\mathbf{z}}\cdot\mathbf{E}_0)\mathbf{k} - (\hat{\mathbf{z}}\cdot\mathbf{k})\mathbf{E}_0\right] \tag{S.14}$$

Using the identity $(\omega\mu_0)^{-1} = Y_0/k_0$, where $k_0$ is the free space wave vector and $Y_0 = \sqrt{\epsilon_0/\mu_0}$ the vacuum admittance, and substituting Eq. (S.13) into Eq. (S.14) we obtain

$$\left[k_0 k_z \bar{\sigma} + 2(\mathbf{k}_{\|}\mathbf{k}_{\|} + k_z^2 \bar{I}_{\|})\right]\cdot\mathbf{E}_{0,\|} = 0, \tag{S.15}$$

which in matrix form is rewritten as

$$\begin{pmatrix} k_0 k_z \sigma_{xx} + 2Y_0(k_x^2 + k_z^2) & 2Y_0 k_x k_y + k_0 k_z \sigma_{xy} \\ 2Y_0 k_x k_y + k_0 k_z \sigma_{xy} & k_0 k_z \sigma_{yy} + 2Y_0(k_y^2 + k_z^2) \end{pmatrix} \begin{pmatrix} E_{0,x} \\ E_{0,y} \end{pmatrix} = \hat{Y}\mathbf{e} = 0 \qquad (S.16)$$

Non-trivial solutions for the vector $\mathbf{e} = (E_{0,x}, E_{0,y})^T$ are found provided that the matrix $\hat{Y}$ is non-singular. To enforce this condition, we impose that $\det[\hat{Y}] = 0$, which yields the dispersion equation:

$$k_x^2 \sigma_1 + (k_x \cos\theta - k_y \sin\theta)^2 \sigma_2 - k_0^2(\sigma_1 + \sigma_2) = 2Y_0 k_0 k_z \left(1 + \frac{\sigma_1 \sigma_2 \sin^2\theta}{4Y_0^2}\right) \qquad (S.17)$$

### 1.4 Solution of the dispersion equation in the lossless scenario

For a lossless metasurface (i.e., if $\Gamma_1 = 0 = \Gamma_2$), the secular equation of the system in (S.16) is solved numerically by first applying a rotation $\hat{R}_\phi = \begin{pmatrix} \cos\phi & \sin\phi \\ -\sin\phi & \cos\phi \end{pmatrix}$ to the conductivity tensor $\hat{\sigma}$, then imposing $k_y' = 0$ and finally introducing the in-plane momentum $\mathbf{q}_{||} = (q, 0)$ (Fig. S3(a)) which is defined in the rotated frame of reference $(k_x', k_y')$. In this way, the value of the surface wave in-plane momentum $q$ is found for each in-plane angle $\phi$. With this transformation, the system matrix in (S.16) becomes

$$\hat{Y}_\phi = \begin{pmatrix} k_0 k_z (\sigma_{xx,i})_\phi + 2Y_0(q^2 + k_z^2) & k_0 k_z (\sigma_{xy,i})_\phi \\ k_0 k_z (\sigma_{xy,i})_\phi & k_0 k_z (\sigma_{yy,i})_\phi + 2Y_0 k_z^2 \end{pmatrix}, \qquad (S.18)$$

where $k_z = i\sqrt{q^2 - k_0^2}$ and

$$\begin{cases} (\sigma_{xx,i})_\phi = \sigma_{xx,i} \cos^2\phi - 2\sigma_{xy,i} \cos\phi \sin\phi + \sigma_{yy,i} \sin^2\phi \\ (\sigma_{xy,i})_\phi = \sigma_{xy,i}(\cos^2\phi - \sin^2\phi) + (\sigma_{xx,i} - \sigma_{yy,i}) \cos\phi \sin\phi \\ (\sigma_{yy,i})_\phi = \sigma_{yy,i} \cos^2\phi + 2\sigma_{xy,i} \cos\phi \sin\phi + \sigma_{xx,i} \sin^2\phi \end{cases} \qquad (S.19)$$

Finally, any in-plane wave vector in the initial frame of reference, $\mathbf{k}_{||} = (k_x, k_y)$, is found imposing $\det[\hat{Y}_\phi] = 0$, solving for $q$ and calculating $k_x = q\cos\phi$ and $k_y = q\sin\phi$ (Fig. S3(b-g)).

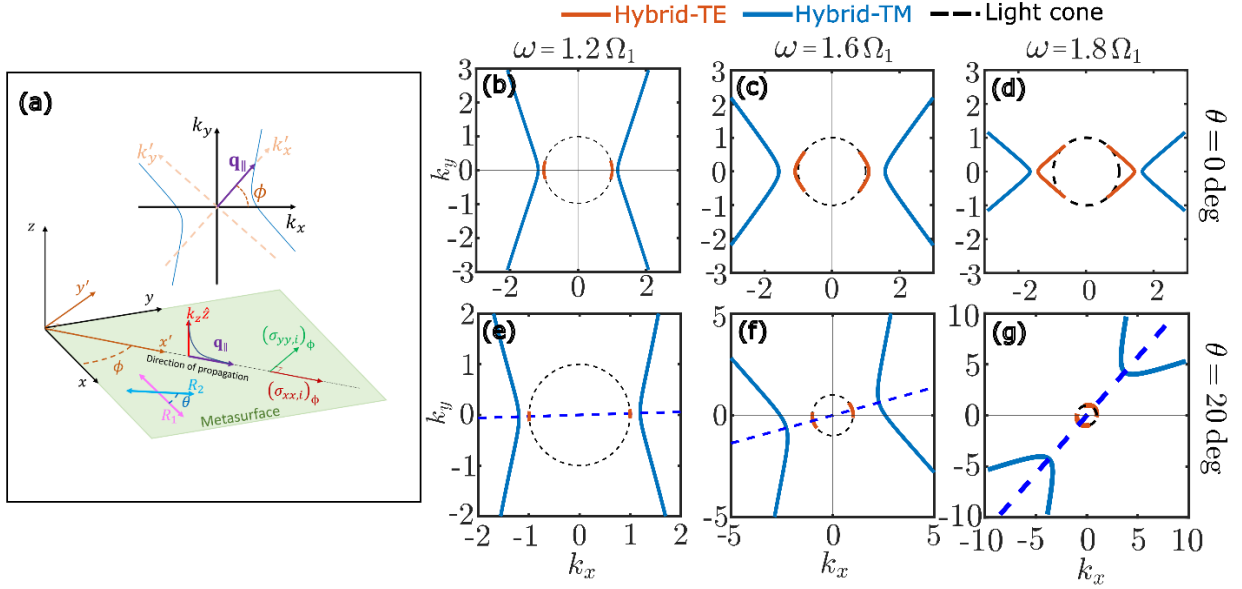

*Figure S3: (a) Sketch of the frame of references and rotation transformations in k-space (top) and real space (bottom) used to calculatae the dispersion equation with lossless resonators. The resulting solutions show that the metasurface supports hybrid-TE (orange) and hybrid-TM (blue) modes in the reststrahlen band for both (b-d) orthogonal resonators and (e-g) twisted resonators, along with a rotation of the iso-frequency contours for both modes.*

## 1.5 Propagation of surface waves in lossy infinitely thin anisotropic metasurfaces

By introducing a non-zero loss rate $\Gamma_1 = \Gamma_2 \neq 0$, we introduce a dissipation channel for the surface wave. Because the background space is lossless, the metasurface is the only source of ohmic losses for propagating surface waves, and since we are interested in strongly confined modes whose dispersion is well below the free space light cone, no radiative losses are considered. We define the complex momentum vector $\mathbf{k} = \mathbf{k}_r + i\mathbf{k}_i$, where the real vectors $\mathbf{k}_r$ and $\mathbf{k}_i$ account for phase propagation direction and decay, respectively, such that the electric field of the surface wave takes the form

$$\mathbf{E} = \mathbf{E}_0 \begin{cases} \exp(-\mathbf{k}_{i,\|} \cdot \mathbf{r}_\| - |k_{i,z}|z), & z \geq 0 \\ \exp(-\mathbf{k}_{i,\|} \cdot \mathbf{r}_\| + |k_{i,z}|z), & z < 0 \end{cases} \quad (S.20)$$

Now that losses are introduced, waves are both evanescently decaying from the metasurface and decaying in plane. We remind the reader that for any plane wave propagating into an anisotropic medium, the direction of propagation of phase and power flow are in general not parallel [36]. Therefore, $\mathbf{k}_r$ and the time averaged Poynting vector, $\mathbf{S} = \mathbf{E} \times \mathbf{H}^*$, are misaligned. This fact holds true for the propagation of surface waves supported by hyperbolic media [16]. Interestingly, for confined surface waves, $\mathbf{k}_i$ and $\mathbf{S}$ must be parallel to correctly account for the direction of dissipation and eventually satisfy the boundary conditions for the electric and magnetic field across the metasurface. This last statement is proved using the integral form of Poynting's theorem, which reads

$$\int_{\partial V} dA\hat{\mathbf{n}} \cdot \mathbf{S} = \int_V dV \frac{1}{2}\Re\left[\mathbf{E}\cdot\mathbf{J}^*\right] \tag{S.21}$$

For the forward propagating modes we are considering, the Poynting vector above the metasurface is $\mathbf{S} = \mathbf{S}_0 e^{-2\mathbf{k}_i\cdot\mathbf{r}}$ and the power loss density is

$$\frac{1}{2}\Re\left[\mathbf{E}\cdot\mathbf{J}^*\right] = \frac{1}{2}\Re\left[\mathbf{E}\cdot(\hat{\sigma}\mathbf{E})^*\right]\delta(z) = \frac{1}{2}\Re\left[\mathbf{E}_0\cdot\mathbf{J}_0^*\right]e^{-2\mathbf{k}_i\cdot\mathbf{r}}\delta(z) = p_L e^{-2\mathbf{k}_i\cdot\mathbf{r}}\delta(z), \tag{S.22}$$

where $p_L = \frac{1}{2}\Re\left[\mathbf{E}_0\cdot\mathbf{J}_0^*\right]$ and $\delta(z)$ accounts for the fact that the metasurface lies on the $(x, y)$ plane.

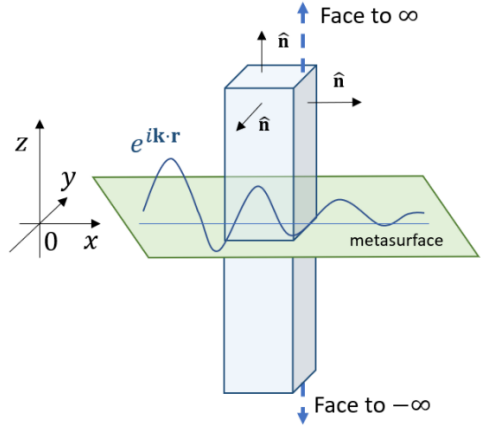

*Figure S4: Sketch of the integration domain used to verify Poynting theorem.*

Consider the parallelepipedal integration domain of volume $V = \Omega_x \times \Omega_y \times \Omega_z = \left[-\frac{L}{2},\frac{L}{2}\right]\times\left[-\frac{L}{2},\frac{L}{2}\right]\times[-h,h]$ in Fig. S4. The flux integral on the left hand side of Eq. (S.21), evaluated in the limit for top and bottom basis of the box to infinity yields

$$\lim_{h\to\infty}\int_{\partial V} dA\hat{\mathbf{n}}\cdot\mathbf{S} = -2\left(\frac{S_x}{k_{i,y}k_{i,z}} + \frac{S_y}{k_{i,x}k_{i,z}}\right)\sinh(k_{i,x}L)\sinh(k_{i,y}L), \tag{S.23}$$

while the volume integral of the power loss density is

$$\lim_{h\to\infty}\int_V dV \frac{1}{2}\Re[\mathbf{E}\cdot\mathbf{J}^*] = \frac{1}{2}\Re[\mathbf{E}_0\cdot\mathbf{J}_0^*]\frac{\sinh(k_{i,x}L)\sinh(k_{i,y}L)}{k_{i,x}k_{i,y}} \tag{S.24}$$

Equating Eq. (S.23) with Eq. (S.24) and rearranging, we obtain

$$\mathbf{k}_{i,\|}\cdot\mathbf{S}_\| = -\frac{k_{z,i}}{2}p_L \tag{S.25}$$

The right-hand side of the last equation accounts for the entire ohmic loss via $p_L$. On the other hand $\mathbf{k}_{i,\|}$ accounts for in-plane decay of the modes, which is attributed to the presence of losses (recall that $\mathbf{k}_{i,\|}=0$ in the lossless scenario). Therefore, since the metasurface is the only dissipation channel, if $\mathbf{k}_{i,\|}$ and $\mathbf{S}$ were not parallel there would be an extra loss contribution which Poynting theorem would be not accounting for. Since we excluded the possibility of dissipation via radiation, the conclusion is that $\mathbf{k}_{i,\|}$ and $\mathbf{S}$ should be parallel.

Interestingly, for a surface mode we can relate the mode damping factor $\eta = |\mathbf{k}_{i,\|}|/|\mathbf{k}_{r,\|}|$ to the power loss density. To do so, we notice that $\mathbf{k}_{i,\|}\cdot\mathbf{S} = |\mathbf{k}_{i,\|}||\mathbf{S}_\||$ and divide both sides of Eq. (S.25) by $|\mathbf{k}_{r,\|}|$, obtaining the relation

$$\eta = -\frac{k_{z,i}}{2|\mathbf{k}_{r,\|}|}\frac{p_L}{|\mathbf{S}_\||} \tag{S.26}$$

which provides a proportionality relation between the damping factor of the surface wave and its power loss normalized by the total real power carried in-plane, i.e. $p_L/|\mathbf{S}_\||$. Finally, for very large $|\mathbf{k}_{r,\|}|$, we obtain the linear relation.

$$\eta = \frac{1}{2}\frac{p_L}{|\mathbf{S}_\||} \tag{S.27}$$

### 1.6 Solution of the dispersion equation in the lossy scenario

To solve for the lossy case, we assume that the loss rates $\Gamma_1 = \Gamma_2 \neq 0$ are small enough to change negligibly the iso-frequency contours, and only affect the broadening of the modes close to the light-cone. Under these assumptions, we use the iso-frequency contours calculated in the lossless scenario, defined by the curve

$\vec{q}_0 = (q_{r,0}\cos\phi, q_{r,0}\sin\phi)$, as a starting guess to calculate the iso-frequency contours in the dissipative case. If loss is turned on, we define the in-plane wave vector in the rotated basis as shown in Fig. S5:

$$\mathbf{q}(\phi) = (q_r + iq_i, iv_i) = \mathbf{q}_r + i\mathbf{q}_i, \tag{S.28}$$

where $\mathbf{q}_r = (q_r, 0)$ is aligned with the direction set by $\phi$, along which phase propagate. Using our theory, the direction of $\mathbf{q}_i = i(q_i, v_i)$ is therefore the same as the one of the group velocity. For hyperbolic media, the group velocity $\mathbf{v}_g$ is orthogonal to the IFCs (Fig. 5(a)). We assume the direction of $\mathbf{v}_g$ is invariant after the loss is considered

$$\mathbf{v}_g = \left( -\frac{d(q_{r,0}(\phi)\cos\phi)}{d\phi}, -\frac{d(q_{r,0}(\phi)\sin\phi)}{d\phi} \right) = (v_x, v_y) \tag{S.29}$$

which in this case is a perturbative approximation in case of weak losses. The derivative above is calculated numerically after the solutions of the lossless case $q_{r,0}(\phi)$ are obtained using (S.18). After normalizing $\mathbf{v}_g$ to have the unit vector $\hat{\mathbf{v}}_g = \mathbf{v}_g / |\mathbf{v}_g| = (\hat{v}_x, \hat{v}_y)$ and rotating the basis $\phi$ we obtain the unit vector

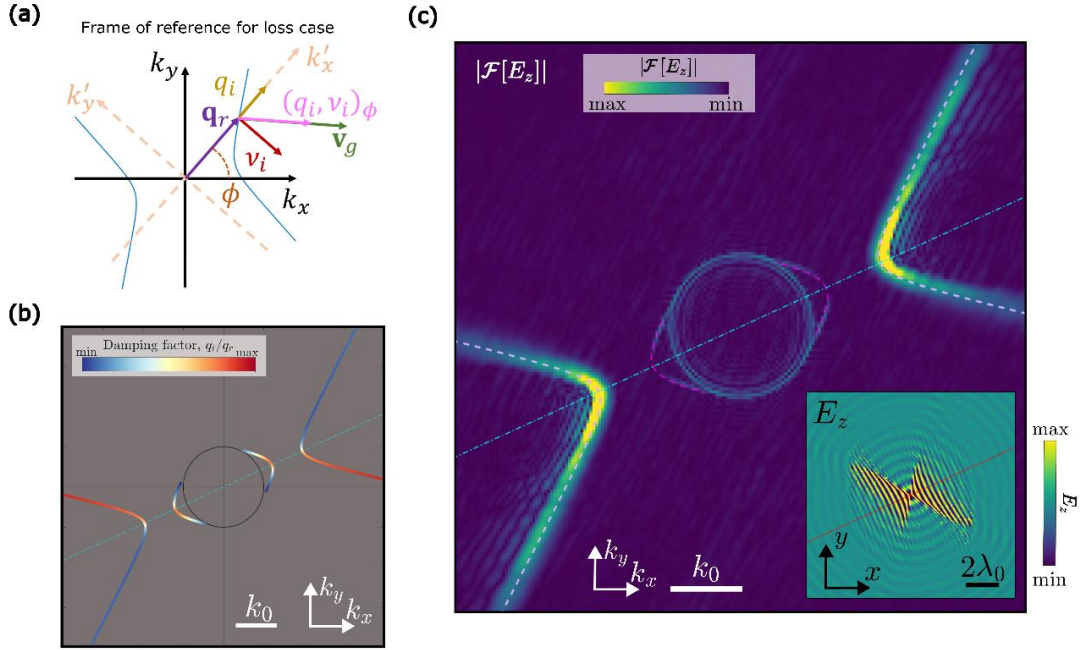

Figure S5: (a) Frame of reference with corresponding rotation transformation used to solve the dispersion equation with lossy resonators. (b) The resulting isofrequency contours show asymmetric damping factor (blue to red colorbar) for both hybrid-TE (closed lenticular contours) and hybrid-TM modes (open hyoerbolic contours). The loss distribution is different for the two modes. The hybrid-TE mode shows larger loss for momenta close to the symmetry axis, whereas the hybrid-TM modes show an ubrapt change of loss close to the symmetry axis, with damping factor saturating for momenta tending to infinity. (c) Fourier transform, $|\mathcal{F}(E_z)|$ of the electric field (inset) generated by a z-oriented point dipole placed above the metasurface. The asymmetric loss reflects in a symmetry breaking of modes lifetime, in agreement with the damping factor in panel (b). The approximation used to solve the dispersion equation with losses (purple and magenta dashed lines) matches with the IFCs obtained via simulations (colormap).

$$\hat{\mathbf{v}}_{g,\phi} = \left(\hat{v}_x \cos\phi + \hat{v}_y \sin\phi, -\hat{v}_x \sin\phi + \hat{v}_y \cos\phi\right) \tag{S.30}$$

The two vectors $\mathbf{q}_i$ and $\hat{\mathbf{v}}_{g,\phi}$ are parallel if

$$\begin{aligned} \frac{q_i}{\sqrt{q_i + v_i}} &= \hat{v}_x \cos\phi + \hat{v}_y \sin\phi \\ \frac{v_i}{\sqrt{q_i + v_i}} &= -\hat{v}_x \sin\phi + \hat{v}_y \cos\phi \end{aligned} \tag{S.31}$$

Therefore, calculating the ratio $v_i / q_i = \beta(\phi)$ we obtain

$$\beta(\phi) = \frac{-\hat{v}_x \sin\phi + \hat{v}_y \cos\phi}{\hat{v}_x \cos\phi + \hat{v}_y \sin\phi}, \tag{S.32}$$

so that the in-plane wave vector becomes

$$\mathbf{q}(\phi) = (q_r + iq_i, i\beta(\phi)q_i) \tag{S.33}$$

In this way we have fixed uniquely our ansatz for the decay direction. Therefore, $\hat{Y}_\phi$ in Eq. (S.16) becomes:

$$\hat{Y}_\phi = \begin{pmatrix} k_0 k_z (\sigma_{xx})_\phi + 2Y_0 \left[(q_r + iq_i)^2 + k_z^2\right] & i2Y_0(q_r + iq_i)\beta q_i + k_0 k_z (\sigma_{xy})_\phi \\ i2Y_0(q_r + iq_i)\beta q_i + k_0 k_z (\sigma_{xy})_\phi & k_0 k_z (\sigma_{yy})_\phi + 2Y_0(k_z^2 - \beta^2 q_i^2) \end{pmatrix}, \tag{S.34}$$

where $k_z = \sqrt{k_0 - q_r^2 + 2iq_r q_i + q_i^2(1 + \beta^2)}$, and

$$\begin{cases} (\sigma_{xx})_\phi = \sigma_{xx} \cos^2\phi - 2\sigma_{xy} \cos\phi \sin\phi + \sigma_{yy} \sin^2\phi \\ (\sigma_{xy})_\phi = \sigma_{xy} (\cos^2\phi - \sin^2\phi) + (\sigma_{xx} - \sigma_{yy}) \cos\phi \sin\phi \\ (\sigma_{yy})_\phi = \sigma_{yy} \cos^2\phi + 2\sigma_{xy} \cos\phi \sin\phi + \sigma_{xx} \sin^2\phi \end{cases} \tag{S.35}$$

After solving the system

$$\begin{cases} \Re\left[\det\left[\hat{Y}_\phi(q_r, q_i)\right]\right] = 0 \\ \Im\left[\det\left[\hat{Y}_\phi(q_r, q_i)\right]\right] = 0 \end{cases} \tag{S.36}$$

for $q_r$ and $q_i$, and finding $v_i = \beta q_i$, the in-plane wave vector, and then the IFC (Fig. S5(b)), of the mode is found as

$$\mathbf{k}_\| = \left((q_r + iq_i)\cos\phi - iv_i \sin\phi, (q_r + iq_i)\sin\phi + iv_i \cos\phi\right) \tag{S.37}$$

In addition, we quantify the loss experienced by the mode with the in-plane damping factor (colormap in Fig. S5(b)), defined as

$$\eta = \frac{|\mathbf{q}_i|}{|\mathbf{q}_r|} = \frac{\sqrt{q_i^2 + v_i^2}}{q_r}, \tag{S.38}$$

which quantifies the decay rate per wavelength for the surface mode. The analytic IFCs obtained solving the determinant of Eq. (S.34) with the proposed approximation (purple and magenta dashed lines in Fig. S5(c)) match the result obtained with full wave simulations of the metasurface.

## 2 Axial dispersion and related features

### 2.1 Derivation of the rotation angle

The goal is to calculatee the rotation angle of the IFCs found using the dispersion equation in Eq. (S.17). Consider a lossless scenario, so that the conductivity tensor is $\hat{\sigma} = i\hat{\sigma}_i$ where $\hat{\sigma}_i \in \mathbb{R}^{2\times 2}$ and the in-plane momentum components of surface modes are real. Let $k_x \to \infty$ and $k_y \to \infty$, and ignore first and zeroth order contributions, then the dispersion equation reduces to:

$$k_x^2 \sigma_{i,1} + (k_x \cos\theta - k_y \sin\theta)^2 \sigma_{i,2} \approx 0. \tag{S.39}$$

Solving for $k_y$ we obtain

$$(k_y)_\pm = k_x \left[ \cot(\theta) \pm \csc(\theta)\sqrt{-\sigma_{i,1}/\sigma_{i,2}} \right], \tag{S.40}$$

which correspond to the coordinates of the two asymptotes of the IFCs upon a induced rotation. The two asymptotes are oriented along the directions set by the angles

$$\tan^{-1}\left(\frac{k_y}{k_x}\right)_\pm = \tan^{-1}\left[ \cot(\theta) \pm \csc(\theta)\sqrt{-\sigma_{i,1}/\sigma_{i,2}} \right]. \tag{S.41}$$

Therefore, the angle subtended by the bisector between the two lines, which correspond to the rotation angle $\gamma(\omega,\theta)$, is:

$$\gamma(\omega,\theta) = \frac{1}{2}\left[ \tan^{-1}(k_y/k_x)_+ + \tan^{-1}(k_y/k_x)_- \right]. \tag{S.42}$$

Finally, using the trigonometric equality $\arctan(a) + \arctan(b) = \arctan\left[(a+b)/(1-ab)\right]$ and after some algebraic manipulations, we obtain the rotation angle

$$\gamma(\omega,\theta) = -\frac{1}{2}\tan^{-1}\left(\frac{\sigma_{i,2}\sin 2\theta}{\sigma_{i,1} + \sigma_{i,2}\cos(2\theta)}\right). \tag{S.43}$$

whose graph versus theta for selected values of frequency is reported in Fig. S6(a).

## 2.2 Stationary point for the rotation angle

The goal is to analyze the dependency of $\gamma(\omega,\theta)$ with respect to the angle $\theta$. To do so, we maximize Eq. (S.43) calculating

$$\frac{\partial \gamma(\omega,\theta)}{\partial \theta} = \frac{\sigma_{i,2}\left(\sigma_{i,2} + \sigma_{i,1}\cos(2\theta)\right)}{\sigma_{i,1}^2 + 2\sigma_{i,1}\sigma_{i,2}\cos(2\theta) + \sigma_{i,2}^2} \tag{S.44}$$

and solving for $\left[\partial \gamma(\omega,\theta)/\partial \theta\right] = 0$ in the interval $0 \leq \theta \leq \pi/2$. Below the critical frequency $\omega^*$ (Fig. S6(b)), where $\sigma_{i,1} > 0$, $\sigma_{i,2} < 0$ and $\sigma_{i,1} > |\sigma_{i,2}|$, a local maximum of $\gamma(\omega,\theta)$ is found for the angle

$$\theta_{\max} = \frac{1}{2}\cos^{-1}\left(-\frac{\sigma_{i,2}}{\sigma_{i,1}}\right), \quad \Omega_1 < \omega < \omega^*, \tag{S.45}$$

which gives the frequency dependent rotation angle maximum:

$$\gamma(\omega,\theta_{\max}) = \gamma_{\max} = -\frac{1}{2}\tan^{-1}\left(\frac{\sigma_{i,2}}{\sqrt{\sigma_{i,1}^2 - \sigma_{i,2}^2}}\right), \quad \begin{cases}\Omega_1 < \omega < \omega^* \\ 0 \leq \theta \leq \pi/2\end{cases}. \tag{S.46}$$

For the case $\omega = \omega^*$ (Fig. S6(c)), we have $\sigma_{i,1} = -\sigma_{i,2}$ and the rotation angle reduces to the line:

$$\gamma(\omega^*,\theta) = \frac{1}{2}\tan^{-1}\left(\cot(\theta)\right), \tag{S.47}$$

which is a monotonically linearly decreasing function of $\theta$, bound between $\gamma(\omega^*,0) = \pi/4$ and $\gamma(\omega^*,\pi/2) = 0$. Finally, if $\omega^* < \omega \leq \omega^*$ (Fig. S6(d)) we have $\sigma_{i,1} < |\sigma_{i,2}|$, and Eq. (S.43) is non-linearly decreasing with respect to $\theta$, bound between the absolute maximum $\gamma(\omega,0) = \pi/4$ and the absolute minimum $\gamma(\omega,\pi/2) = 0$.

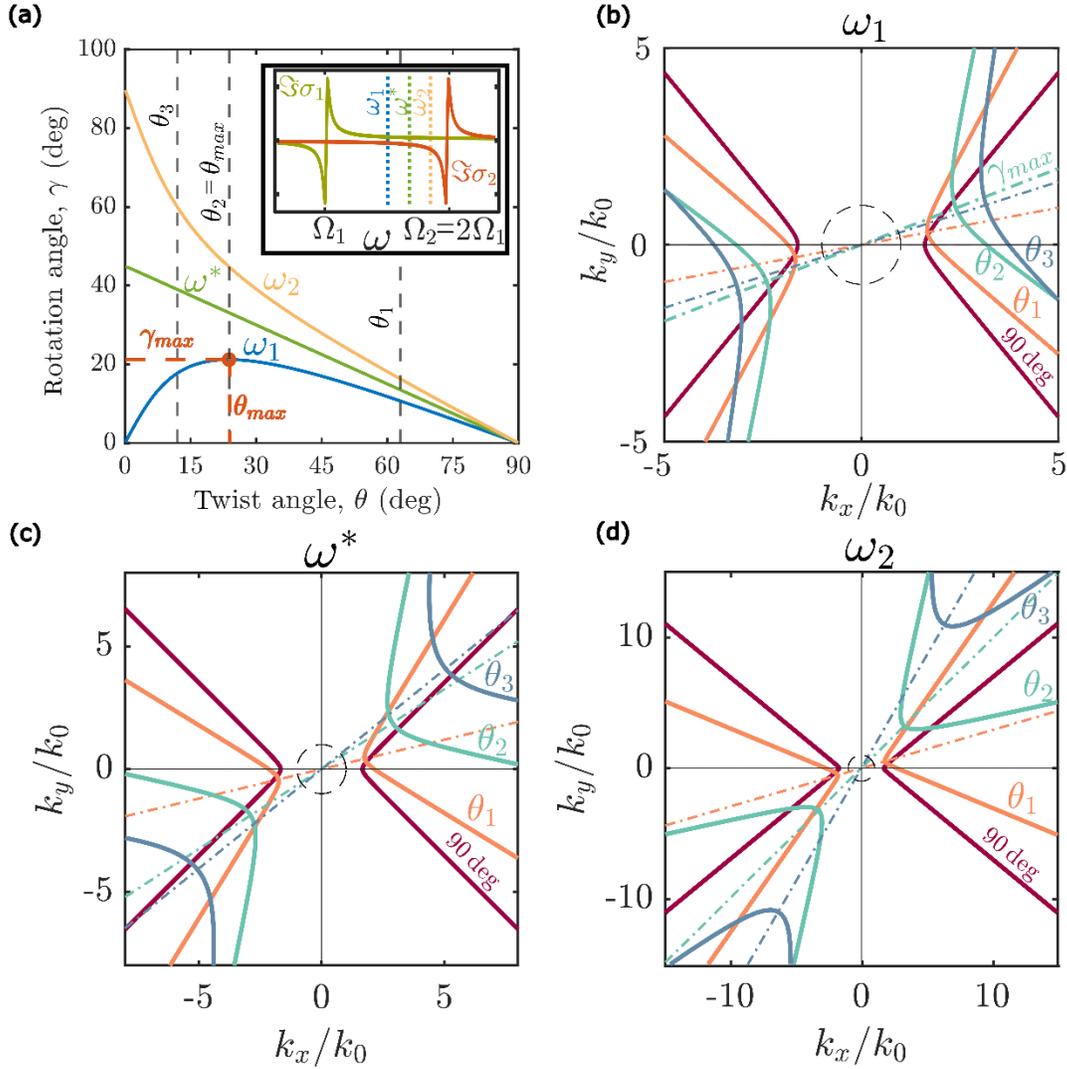

*Figure S6 (a) The rotation angle is found in three phases depending on the working frequency. Here we consider three frequencies across the critical frequency (inset). Below the critical frequency (blue line) $\gamma$ grows as the tiwst angle reduces, reaching a maximum $\gamma = \gamma_{max}$ for $\theta = \theta_{max}$ and eventually turning back to zero. If the frequency equals $\omega^*$ (green line), the rotation angle grows linearly as $\theta$ reduces, reaching its maximum $\gamma = 45°$ at $\theta = 0$. Above $\omega^*$ the rotation angle grows steadily to $90°$ as the angle reduces. For the chosen angles $[\theta_1, \theta_2, \theta_3] = [65°, 23.3°, 12°]$ the isofrequency contours rotate in k space (b) turnin back if $\omega < \omega^*$, or increasing if (c) $\omega = \omega^*$ or (d) $\omega > \omega^*$, in contrast with the case of orthogonal resonators (red) where the contours never rotate.*

## 2.3 Controllable mode confinement

In our letter, we have shown how modes get confined upon a rotation of the resonators. The goal of this section is quantifying the confinement for both the hybrid-TM and the hybrid-TE modes supported by the metasurface, as we change $\theta$. Without loss of generality, we carry out the analysis assuming a lossless metasurface, i.e., setting $\Gamma_1 = 0 = \Gamma_2$. For convenience, we work in a basis rotated by $\gamma(\omega, \theta)$, such that the conductivity tensor is always diagonal as presented in the previous section, i.e.:

$$\hat{\sigma}' = \hat{R}_\gamma \hat{\sigma} \hat{R}_\gamma^T = i\hat{\sigma}_i' = i \begin{pmatrix} \sigma_{i,1}' & 0 \\ 0 & \sigma_{i,2}' \end{pmatrix}, \tag{S.48}$$

In this new basis, the coordinate of the $k$-space are $\left(k_x', k_y'\right)$ and the IFCs are symmetric with respect to the $k_x'$ axis. To find the vertex of both hybrid-TM and the hybrid-TE modes we look for solutions at $k_y' = 0$. In this rotated frame, the dispersion equation reduces to:

$$\sigma_{i,1}'(k_x')^2 - \left(\sigma_{i,1}' + \sigma_{i,2}'\right)k_0^2 = -\frac{1}{2}ik_0 Y_0 \sqrt{k_0^2 - (k_x')^2}\left(4 - \frac{\sigma_{i,1}' \sigma_{i,2}'}{Y_0^2}\right). \tag{S.49}$$

Squaring both sides and rearranging we obtain the factorized quartic equation in $k_x'$

$$\left[(k_x')^2(\sigma_{i,1}')^2 - k_0^2\left(4Y_0^2 + (\sigma_{i,1}')^2\right)\right]\left[4(k_x')^2 Y_0^2 - k_0^2\left(4Y_0^2 + (\sigma_{i,2}')^2\right)\right] = 0, \tag{S.50}$$

whose solutions, in terms of the non-rotated conductivity tensor, are

$$q_{v,1}(\omega, \theta) = \pm k_0 \sqrt{1 + \left(\frac{4Y_0}{\sigma_{i,1} + \sigma_{i,2} + \sqrt{\sigma_{i,1}^2 + \sigma_{i,2}^2 + 2\sigma_{i,1}\sigma_{i,2}\cos(2\theta)}}\right)^2} \tag{S.51}$$

and

$$q_{v,2}(\omega, \theta) = \pm k_0 \sqrt{1 + \left(\frac{\sigma_{i,1} + \sigma_{i,2} - \sqrt{\sigma_{i,1}^2 + \sigma_{i,2}^2 + 2\sigma_{i,1}\sigma_{i,2}\cos(2\theta)}}{4Y_0}\right)^2}, \tag{S.52}$$

which corresponds to the coordinate of the vertices of the IFCs for hybrid-TM modes (least confined point) and hybrid-TE modes (most confined point), respectively. Notice that the frequency dependency is hidden in the reactive strength of the two physical resonator sets, i.e., $\sigma_{i,1}$ and $\sigma_{i,2}$. Plots of $q_{v,2}(\omega, \theta)$ dependency with respect to $\omega$ and $\theta$ are reported in Fig. S7(a) and Fig. S7(b), respectively. Focusing on the quasi-TM branch, as $\theta$ is reduced, $q_{v,2}$ increases monotonically until the frequency reaches a threshold. Above this value [Fig. S7(b)], for decreasing angles $\theta$, modes are first translated closer to the light cone, and then pushed to larger momenta upon reaching a minimum distance at the cusp.

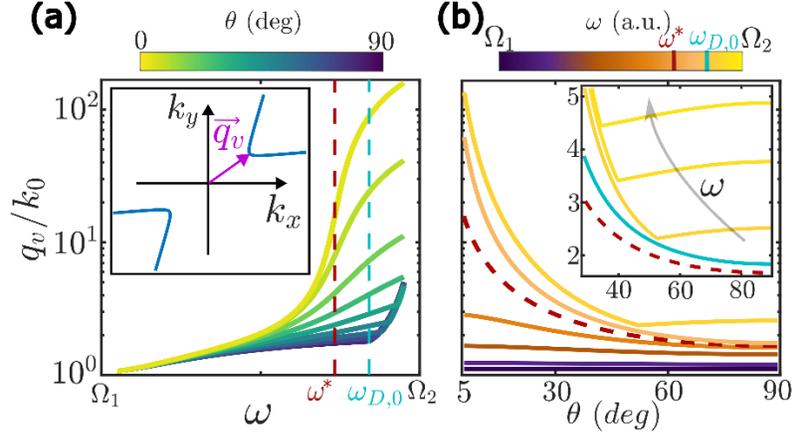

*Figure S7(a) The minimal wavector of q-TM waves supported by the hyperbolic IFCc from the origin $q_v$, (purple arrow in the inset) monotonically increases with frequency. $q_v$ is insensitive to twisting near the lower resonance $\Omega_1$, acquiring a strong tunability for $\omega \gtrsim \omega^*$, after which the contours shift dramatically away from the light cone, reaching a maximum near the upper resonance $\Omega_2$. (b) Reducing $\theta$ increases $q_v$ for $\omega < \omega_{D,0}$ (light-blue dotted line) above which a kink occours at the angles $\theta_D(\omega)$ (inset) due of an accidental band degeneracy.*

To study this behavior, we derive the change of the two vertices in Eq. (S.51) and Eq. (S.52) with respect to the angle $\theta$:

$$\frac{\partial q_{v,1}(\omega,\theta)}{\partial \theta} = \frac{16 k_0 \sqrt{2} \sigma_{i,1} \sigma_{i,2} \sin(2\theta)}{\alpha(\sigma_{i,1}+\sigma_{i,2}+\alpha)^2 \sqrt{8Y_0^2 + \sigma_{i,1}^2 + \sigma_{i,2}^2 + 2\sigma_{i,1}\sigma_{i,2}\cos^2(\theta) + (\sigma_{i,1}+\sigma_{i,2})\alpha}} < 0 \qquad (S.53)$$

and

$$\frac{\partial q_{v,2}(\omega,\theta)}{\partial \theta} = \frac{k_0 \sigma_{i,1}\sigma_{i,2}(\sigma_{i,1}+\sigma_{i,2}-\alpha)\sin(2\theta)}{8\alpha \sqrt{Y_0^2 + \frac{1}{16}(\sigma_{i,1}+\sigma_{i,2}-\alpha)^2}} > 0, \qquad (S.54)$$

respectively, where $\alpha = \sqrt{\sigma_{i,1}^2 + \sigma_{i,2}^2 + 2\sigma_{i,1}\sigma_{i,2}\cos(2\theta)}$.

The two derivatives maintain opposite signs in the entire hyperbolic frequency range suggest that a crossing point of the hybrid-TE and hybrid-TM IFCs exists for a certain value of $\theta$, where bands touch [18], and the hybrid-TM mode bands reaches its minimum distance from the light cone. Intriguingly, this degeneracy is conserved even in the monoclinic scenario, and in a shear metasurface its occurrence in momentum space becomes tunable. In fact, band degeneracy is achieved choosing the angle $\theta_D(\omega)$ (found equating Eq. (S.51) with Eq. (S.52)):

$$\theta_D(\omega) = \frac{1}{2}\cos^{-1}\left(\frac{\sigma_{i,1}\sigma_{i,2} + 8Y_0^2}{\sigma_{i,1}\sigma_{i,2}}\right). \tag{S.55}$$

The minimal distance of the vertex at the degeneracy point normalized to $k$ is found substituting Eq. (S.55) in (S.52), having

$$\frac{q_{v,D}(\omega)}{k_0} = \sqrt{1 + \left(\frac{\sigma_{i,1} + \sigma_{i,2}}{4Y_0} - \sqrt{1 + \left(\frac{\sigma_{i,1} + \sigma_{i,2}}{4Y_0}\right)^2}\right)^2}, \tag{S.56}$$

The in-plane direction at which we find $q_{v,D}(\omega)$ is found substituting Eq. (S.55) in (S.43) and is:

$$\phi_D = -\frac{1}{2}\tan^{-1}\left[\frac{\sqrt{-8Y_0^2(2\sigma_{i,1}\sigma_{i,2} + 8Y_0^2)}}{\sigma_{i,1}\sigma_{i,2}\left(\sigma_{i,1}^2 + \sigma_{i,1}\sigma_{i,2} + 8Y_0^2\right)}\right]. \tag{S.57}$$

# 3 Quantification of the shear effect

## 3.1 Defining the effective shear in the conductivity tensor, $S_\sigma$

Consider the lossless metasurface described by the complex conductivity tensor

$$\hat{\sigma} = i\hat{\sigma}_i = i\begin{pmatrix} \sigma_{i,xx} & \sigma_{i,xy} \\ \sigma_{i,xy} & \sigma_{i,yy} \end{pmatrix}. \tag{S.58}$$

The matrix $\hat{\sigma}$ is purely imaginary, and therefore it can be diagonalized by a unitary transformation $\hat{R}_\gamma$ as

$$\hat{\sigma}' = \hat{R}_\gamma \hat{\sigma} \hat{R}_\gamma^T = i\begin{pmatrix} \sigma_{i,xx}' & 0 \\ 0 & \sigma_{i,yy}' \end{pmatrix} \tag{S.59}$$

where $\sigma_{i,xx}'$ and $\sigma_{i,yy}'$ are the eigenvalues of $\hat{\sigma}_i$, and $\hat{R}_\gamma = \begin{pmatrix} \cos\gamma & \sin\gamma \\ -\sin\gamma & \cos\gamma \end{pmatrix}$ is a real rotation matrix, with rotation angle $\gamma$. In a physical picture, the eigenvectors of $\hat{\sigma}'$ describe the directions along which a the electric field of a propagating evanescent mode $\mathbf{E} = \mathbf{E}_0 e^{i\mathbf{k}_\parallel \cdot \mathbf{r}_\parallel} e^{-k_z z}$ would induce two orthogonal polarization currents $\mathbf{J}_i = \hat{\sigma}\mathbf{E}$ on the metasurface. The quantity $\gamma$ is found calculating the angle that subtends between one of the eigenvectors of $\hat{\sigma}_i$ and the $x$ axis as

$$\gamma = \frac{1}{2}\tan^{-1}\left(\frac{2\sigma_{i,xy}}{\sigma_{i,xx} - \sigma_{i,yy}}\right) \tag{S.60}$$

This relation, agrees with Eq. (S.43), where we derived $\gamma(\omega,\theta)$ starting for the dispersion of the surface modes supported by the metasurface, also agreeing the result proposed in [REF elasticity].

Introducing losses, the conductivity tensor becomes

$$\hat{\sigma} = \hat{\sigma}_r + i\hat{\sigma}_i = \begin{pmatrix} \sigma_{r,xx} & \sigma_{r,xy} \\ \sigma_{r,xy} & \sigma_{r,yy} \end{pmatrix} + i\begin{pmatrix} \sigma_{i,xx} & \sigma_{i,xy} \\ \sigma_{i,xy} & \sigma_{i,yy} \end{pmatrix}, \tag{S.61}$$

where $\hat{\sigma}_r = \begin{pmatrix} \sigma_{r,xx} & \sigma_{r,xy} \\ \sigma_{r,xy} & \sigma_{r,yy} \end{pmatrix}$ describes the ohmic dissipations. The matrix $\hat{\sigma}_r$ is not diagonalized by the rotation real rotation $\hat{R}_\gamma$ because $\hat{\sigma}_r$ and $\hat{\sigma}_i$ do not share the same eigenvectors, therefore in the rotate basis we obtain:

$$\hat{\sigma}' = \hat{R}_\gamma (\hat{\sigma}_r + i\hat{\sigma}_i)\hat{R}_\gamma^T = \begin{pmatrix} \sigma_{r,xx}' & \sigma_{r,xy}' \\ \sigma_{r,xy}' & \sigma_{r,yy}' \end{pmatrix} + i\begin{pmatrix} \sigma_{i,xx}' & 0 \\ 0 & \sigma_{i,yy}' \end{pmatrix}, \tag{S.62}$$

where

$$\begin{cases} \sigma_{r,xx}' = \sigma_{r,xx}\cos^2\gamma - 2\sigma_{r,xy}\cos\gamma\sin\gamma + \sigma_{r,yy}\sin^2\gamma \\ \sigma_{r,xy}' = \sigma_{r,xy}(\cos^2\gamma - \sin^2\gamma) + (\sigma_{r,xx} - \sigma_{r,yy})\cos\gamma\sin\gamma \\ \sigma_{r,yy}' = \sigma_{r,yy}\cos^2\gamma + 2\sigma_{r,xy}\cos\gamma\sin\gamma + \sigma_{r,xx}\sin^2\gamma \end{cases} \tag{S.63}$$

Then, the resulting matrix $\hat{\sigma}'$ is non-diagonal, and the off-diagonal elements $\gamma$ are purely real. These elements are responsible for the loss symmetry breaking in the system, and we may use them to define the shear factor

$$\hat{\sigma}_r' = \begin{pmatrix} \sigma_{r,xx}' & \sigma_{r,xy}' \\ \sigma_{r,xy}' & \sigma_{r,yy}' \end{pmatrix}. \tag{S.64}$$

## 3.2 Defining the effective shear in the iso-frequency contours, $S_\sigma$

The figure of merit $S_\sigma$ is a global parameter, agnostic of the dispersion of the surface wave guided by the conductive sheet. We therefore introduce the figure of merit

$$S_\eta(\omega,\theta) = \frac{\eta_+ - \eta_-}{\eta_+ + \eta_-} = \frac{\left(\frac{|q_i|}{|q_r|}\right)_+ - \left(\frac{|q_i|}{|q_r|}\right)_-}{\left(\frac{|q_i|}{|q_r|}\right)_+ + \left(\frac{|q_i|}{|q_r|}\right)_-}, \tag{S.65}$$

where $(\eta)_+ = (|q_i|/|q_r|)_+$ and $(\eta)_- = (|q_i|/|q_r|)_-$ are the damping factors for mirror-symmetric points on the IFCs on the bright and dark branch, respectively. The quantity $S_\eta(\omega,\theta)$ measures the loss asymmetry for modes lying on the bright and dark branches in the non-orthogonal system (Fig. S8(a)). The figure of merit is calculated at large momenta, far from the light cone, where the damping factor saturates and the difference $\eta_+ - \eta_-$ becomes constant (Fig. S8(b)) and is proportional to the difference in power loss densities for the two modes (see Eq. (S.27)).

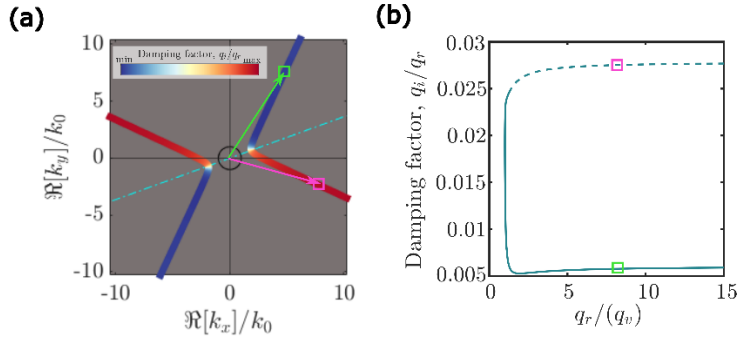

*Figure S8 (a) The figure of merit $S_\eta$ is calculated considering large in-plane momenta (green and pink arrows) for axysimmetric points (green and pink square). (b) The damping factor saturates for large momenta (pink and green square).*

# 4 Further analyses on the Purcell factor

We extend the calculation of the Purcell factor enhancement reported in Fig. 4b of the main text (here reported in Figure S9a for convenience) to the case of a different source position. In our previous analysis we placed the source at a distance $d_e = 1 \text{ mm} = \lambda^*/217$ from the metasurface, where $\lambda^* = 2\pi c/\omega^* = 0.217$ m, and showed global Purcell factor enhancement within the entire hyperbolic bandwidth, peaking at $\omega^*$. We observe that an increase of the source distance to $d_e = 4 \text{ mm} = \lambda^*/54$, does not alter significantly the physical response of the system, as shown if Figure S9b. In fact, an overall Purcell factor enhancement of $\sim 1.8$ is observed across the hyperbolic bandwidth, peaking at $2.3$ times at the critical frequency. The lower enhancement compared to the case of Figure S9a is ascribed to a less efficient coupling of the source to strongly confined surface wave momenta.

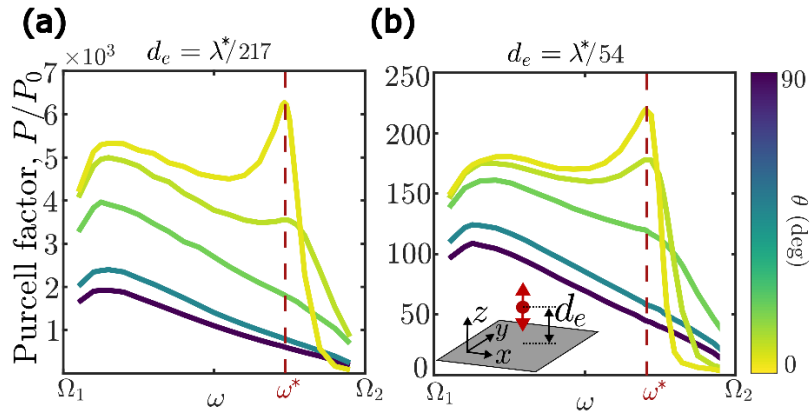

*Figure S9 Purcell factor enhancement for a point source placed at (a) $d_e = \lambda^*/217$ and (b) $d_e = \lambda^*/54$ from the metasurface, where $\lambda^* = 2\pi c/\omega^*$. Different positions of the source do not change the physical property of the system, which shows global spontaneous emission rate within the hyperbolic bandwidth peaking at $\omega^*$. For these calculations, we considered $\Omega_1 = \Omega_2/2 = 5$ GHz, $\gamma_1 = \gamma_2 = 0.02\Omega_1$, $N_1 = 2N_2 = 1$, and $\omega = \omega^* = 1.733\,\Omega_1$, such that $\lambda^* = 0.217$ m.*